\documentclass{article}
\usepackage{spconf,amsmath}
\usepackage{graphicx} %
\usepackage{booktabs}
\usepackage{pgfplots}
\pgfplotsset{compat=1.18}
\usepackage{tabularx} %
\usepackage{makecell} %
\usepackage{enumitem}
\setlist{nosep, leftmargin=14pt}
\usepackage{setspace}
\usepackage{mwe} %
\usepackage{subcaption} %

\title{Quantitative Imaging of $^{55}$Co and $^{18}$F-Labeled Tracers in a Single ``Multiplexed'' PET Imaging Session}
\name{Sarah J. Zou$^{1}$, Irene Lim$^{2}$, Jackson W. Foster$^{1}$, Garry Chinn$^{3}$, Hailey A. Houson$^{4}$,
Suzanne E. Lapi$^{4}$, Jianghong Rao$^{3, 7}$, Craig S. Levin$^{1, 3, 5, 6, 7}$}

\address{
$^1$ Department of Electrical Engineering, Stanford University\\
$^2$ Chemistry and Biochemistry Department, Bates College\\
$^3$ Department of Radiology, Stanford University\\
$^4$ Department of Radiology, University of Alabama at Birmingham\\
$^5$ Department of Bioengineering, Stanford University\\
$^6$ Department of Physics, Stanford University\\
$^7$ Molecular Imaging Program at Stanford (MIPS), Stanford University
}

\begin{document}
\maketitle
\begin{abstract}
In this study, we explore the use of Co-55 as a radioisotope for multiplexed PET (mPET) by utilizing its emission of a prompt gamma-ray in cascade with a positron during decay. We leverage the prompt-gamma signal to generate triple coincidences for a Co-55-labeled tracer, allowing us to distinguish it from a tracer labeled with a pure positron emitter, such as F-18. By employing triple versus double coincidence detection and signal processing methodology, we successfully separate the Co-55 signal from that of F-18. Phantom studies were conducted to establish the correlation between Co-55 double and triple coincidence counts and Co-55 activity. Additionally, we demonstrate the potential for quantifying hot spots within a warm background produced by both Co-55 and F-18 signals in a single PET scan. Finally, we showcase the ability to simultaneously image two tracers \textit{in vivo} in a single PET session with mouse models of cancer.
\end{abstract}
\section{Introduction}
\label{sec:intro}
Positron emission tomography (PET) is a vital and versatile tool for observing and quantifying biological pathways \textit{in vivo} \cite{muehllehner_positron_2006}. Small animal PET imaging is used to guide the development of new therapeutics for diseases such as cancer and neurological disorders by tracking how a radiotracer (a radioactively-tagged molecule of interest) distributes in the body \cite{kuntner_quantitative_2014}. Currently, conventional PET is limited to imaging one tracer at a time. However, simultaneous multi-tracer or multiplexed PET (mPET) would be highly useful due to its ability to visualize and quantify two (and potentially more) biological pathways in one scan session \cite{pratt_simultaneous_2023}. One of the obstacles preventing the realization of mPET is the conventional signal processing used in PET. PET radiotracers are tagged with radioactive isotopes that emit positrons ($\beta^+$) during the decay process. These positrons then annihilate with electrons around them to produce two oppositely-directed 511 keV photons. These pairs of photons are detected by the PET scanner and used to reconstruct the 3-D distribution of the radiotracer \cite{wernick2004emission}. Due to the uniform energy of annihilation photons, reliance on their detection alone is insufficient for differentiating between various radiotracers.

One way to achieve mPET is by imaging one radiotracer tagged with a pure positron-emitting ($\beta^+$) isotope and another tagged with a prompt-gamma+positron-emitting ($\beta^+-\gamma$) isotope \cite{andreyev_dual-isotope_2011, gonzalez_multiplexed_2011}. In this approach, the $\beta^+-\gamma$ isotope releases an additional photon in cascade with positron emission, resulting in a triple coincidence (three photons) that is distinguishable from the conventional double coincidence (two photons) from a pure positron emitter.

In this paper, we further explore the concept of multiplexed PET (mPET), demonstrating for the first time quantitative imaging of Co-55 simultaneously with F-18 (a $\beta^+$ isotope) in both phantom and mice experiments. Co-55 is a $\beta^+-\gamma$ isotope that emits a prompt-gamma of 931 keV in 75\% of positron emissions \cite{mastren_cyclotron_2015}. Further, Co-55 is a long-lived radionuclide that we believe will be useful for labeling and tracking an antibody that targets important biomarkers for cancer immunotherapy.

\subsection{Previous Work}
\label{sec:PreviousWorks}

The first descriptions of using the triple coincidence produced from a $\beta^+-\gamma$ emitter to distinguish from two photon coincidences from a pure $\beta^+$ emitter were presented in \cite{andreyev_dual-isotope_2011, gonzalez_methods_2011, olcott2016dual, olcott2015methods}. Fukuchi \textit{et. al} was the first study to image simultaneous dual-tracer PET in animals by imaging FDG (a pure $\beta^+$ emitter) and Na-22 (a $\beta^+-\gamma$ emitter) \cite{fukuchi_positron_2017}. More recently, Pratt \textit{et. al} showed dual-tracer PET with various combinations of pure $\beta^+$ emitting isotope labeled tracers imaged with $\beta^+-\gamma$ emitting isotope labeled tracers such as Zr-89 with I-124, F-18 with I-124, and Zr-89 with Y-86 \cite{pratt_simultaneous_2023}.

\section{Methods}
\label{sec:methods}
\subsection{PET System Used for mPET Studies}
The Siemens Inveon DPET system (Siemens Healthcare, Knoxville, TN) \cite{bao_performance_2009} was used to explore mPET. Although not designed for acquiring two PET tracers simultaneously, it does provide information enabling us to acquire the data we need.  
The Inveon system registers detected triple coincidences by recording them as pairs of two-detector coincidences sharing a common detector. For instance, a triple coincidence among detectors (a, b, c) is recorded as two-detector coincidences (a, b) and (b, c) \cite{pratt_simultaneous_2023}. At high count rates, these pairs may not appear consecutively; however, the shared detector pattern can typically be identified within two or three recorded coincidences.

During acquisition, the lower energy window was set at 350 keV and the upper energy window was set at 814 keV. The timing window was set at 3.438 ns and scan time length was 15 minutes.

\subsection{Phantom Studies}
\subsubsection{Uniform Decay} \label{sec: uniform_decay}
Ten 2 mL tubes were filled with 1.5 mL of 80 $\mu$Ci of Co-55. The tubes were placed with 2 tubes in radial axis and 5 tubes along axial axis and centered in the field of view. Over the course of three days, as the Co-55 decayed, scans were taken when the activity levels in the tubes were at 80, 70, 60, 50, 40, 30, 20, 10, 7, and 4 $\mu$Ci.

\subsubsection{Micro Hollow Sphere Phantom}
Using the Micro Hollow Sphere phantom (Data Spectrum, Durham, NC), which includes four hollow spheres within a fillable background, we prepared the spheres with the following contents: (1) a 1:1 mixture of F-18 and Co-55, (2) Co-55 only, (3) F-18 only, and (4) water. The phantom body was then filled with a 1:1 mixture of F-18 and Co-55, at one-third the activity concentration of sphere 1, which contained the mixed isotopes.

\subsection{Animal Study}
The focus of this paper is to describe the methodology of tracer signal unmixing needed for mPET. However, we also present \textit{in vivo} studies to demonstrate proof of principle for the unmixing methodology.

Six-week-old female Balb/c mice, sourced from Jackson Labs (Sacramento, CA), were used in the animal experiments. A syngeneic mouse colon cancer line, CT26, was implanted subcutaneously (1 million cells in 0.1 mL of saline) in the right shoulders of the mice 7 days prior to the study. When the tumors reached 100--200 cubic millimeters, probes were prepared and injected retro-orbitally. Co-55 was complexed to a DOTA-aPD-L1 monoclonal antibody conjugate (150 $\mu$Ci/mg specific activity) that binds to the PD-L1 antigen on tumor cells. However, it appears that there may have been some unstable Co-55 antibody complex in the final product, leading to rapid clearance and high signal in the kidneys. Despite this limitation, the experiments were still informative for evaluating the proposed mPET signal unmixing strategies. Due to the rapid clearance of the Co-55-labeled antibody, FDG (80 $\mu$Ci in 0.1 mL) was injected 60 minutes before imaging, and the Co-55-labeled antibody (25 $\mu$Ci in 0.1 mL) was injected 15 minutes before imaging. Single-tracer and dual-tracer mice were prepared, and PET scans were conducted as described above. At the end of the experiments, the mice were sacrificed.

\subsection{Image Reconstruction and Analysis}
The listmode data were divided into double coincidence listmode datasets and triple coincidence listmode datasets. Ordered subset expectation-maximization (OSEM) reconstruction was applied with 70 subsets and 100 iterations for doubles list-mode datasets to create double images, and with 70 subsets and 120 iterations for triples list-mode datasets to create triple images. These reconstruction parameters were selected as they provided the best balance between resolution and noise. The differing settings for double and triple datasets are the result of the the triple dataset being substantially smaller than the double dataset due to lower triple coincidence sensitivity. Both double and triple images were normalized with  respective images from a Co-55 uniform cylinder scan that covered the field-of-view (FOV).

We used the method from \cite{moore_simultaneous_2019} to obtain the F-18 distribution and Co-55 distribution. Co-55 appears in the double images due to the scanner detecting only 2 out of 3 photons in a triple coincidence and the pure positron branch of Co-55 decay. To separate the F-18 distribution from Co-55 distribution,  we applied a scaling factor ($g$) that represents the ratio of detected double coincidences to triple coincidences of Co-55 to obtain the F-18 image (F-18 image = Double Image - g * Triple Image). From image analysis of the Uniform Decay study (Sec. \ref{sec: uniform_decay}), we found that image value of Co-55 doubles were on average 12.6 times the Co-55 triple image values, so we used $g = 12.6$. We also use a ($f$) factor to correct for random triples that come from the pure positron emitter so we use (Co-55 image = Triple Image - f * Double Image) to create prompt-gamma positron emitter image. From a scan of FDG only, we estimated $f = 0.005813$.

For the uniform decay tubes, we placed regions of interest (ROIs) at the same locations on both the double and triple images. The tubes were approximately 10 mm in diameter at their widest point and 40 mm in height. We drew 10 ROIs, each measuring 5 mm $\times$ 5 mm $\times$ 18 mm, in the center of the tubes. To evaluate the accuracy of the F-18 and Co-55 images, ROIs were centered on each of the four spheres in the hollow sphere phantom as cubes with side lengths equal to half sphere diameters. Additionally, a background ROI, measuring 9 mm $\times$ 9 mm $\times$ 20 mm, was placed in an area without spheres.

\section{Results}
\subsection{Phantom Studies}
\subsubsection{Scan of Tubes of Uniform Concentration}
\begin{figure}
    \centering
    \includegraphics[width=\linewidth]{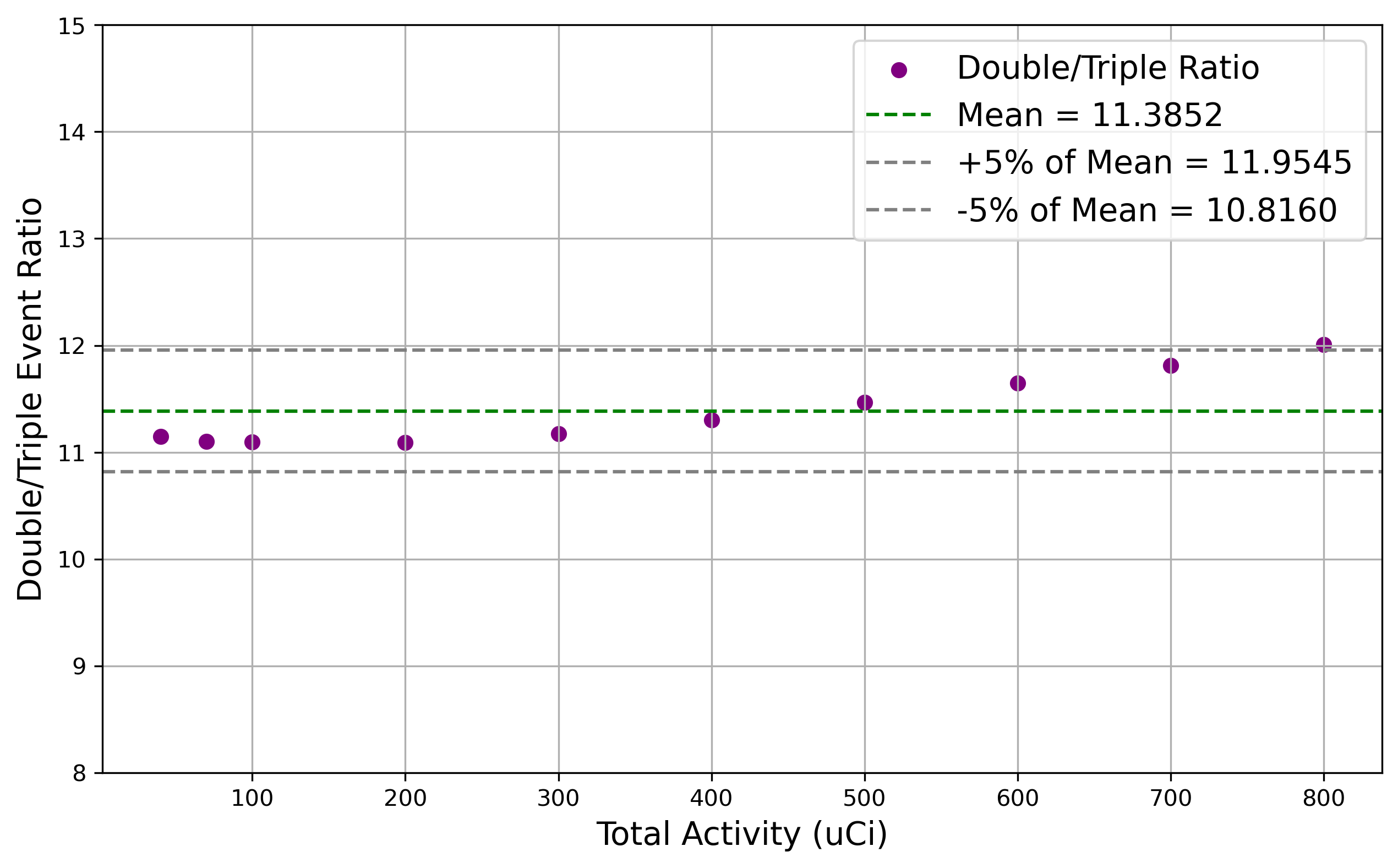}
    \caption{Ratio of double to triple event counts over varying total Co-55 activities in the tubes of uniform activity}
    \label{fig:double_triple_event}
\end{figure}

\begin{figure}[ht]
    \centering
    \begin{minipage}[b]{\linewidth}
        \centering
        \includegraphics[width=\linewidth]{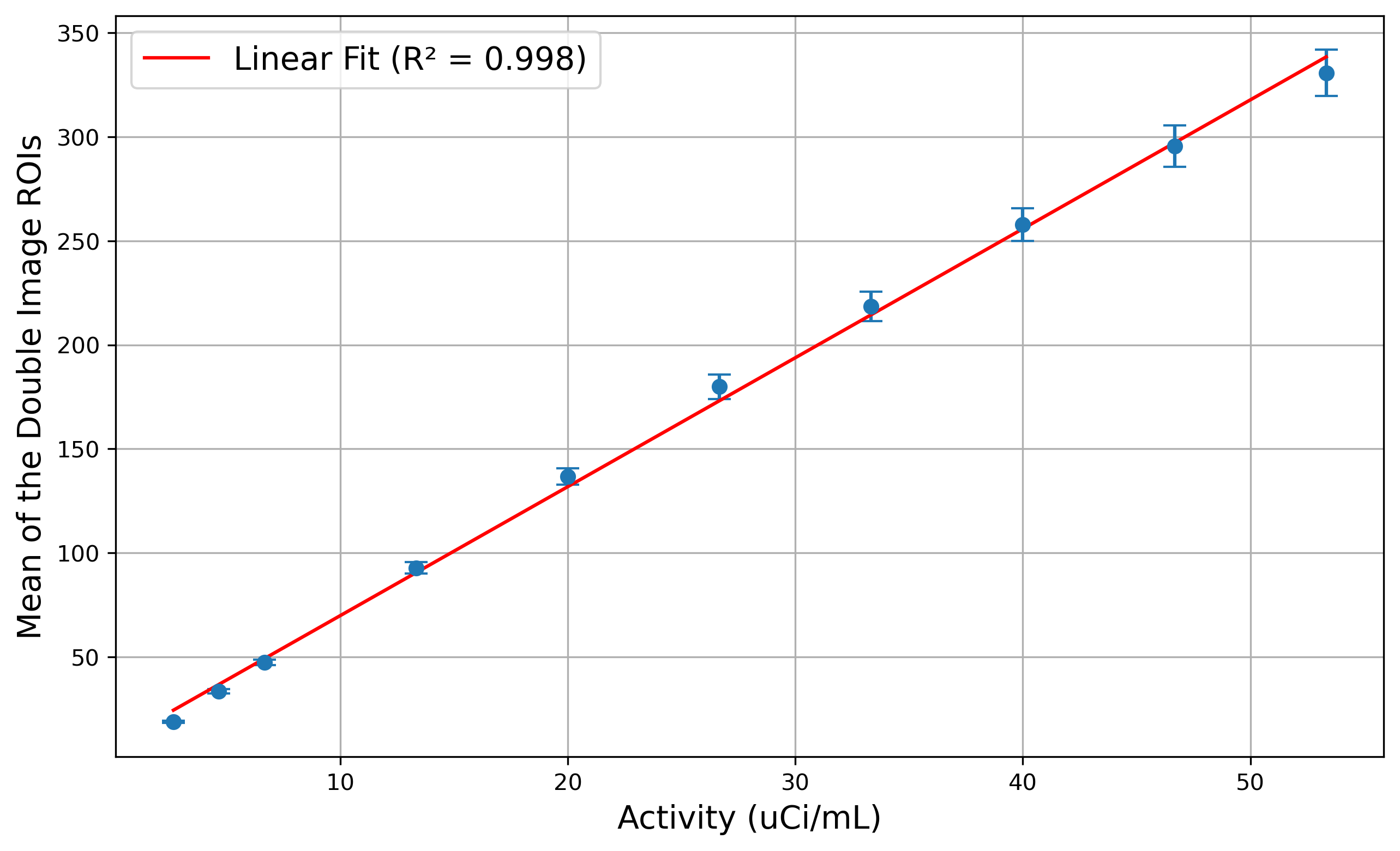}
        \subcaption{}
    \end{minipage}
    \hspace{0.05\linewidth}
    \begin{minipage}[b]{\linewidth}
        \centering
        \includegraphics[width=\linewidth]{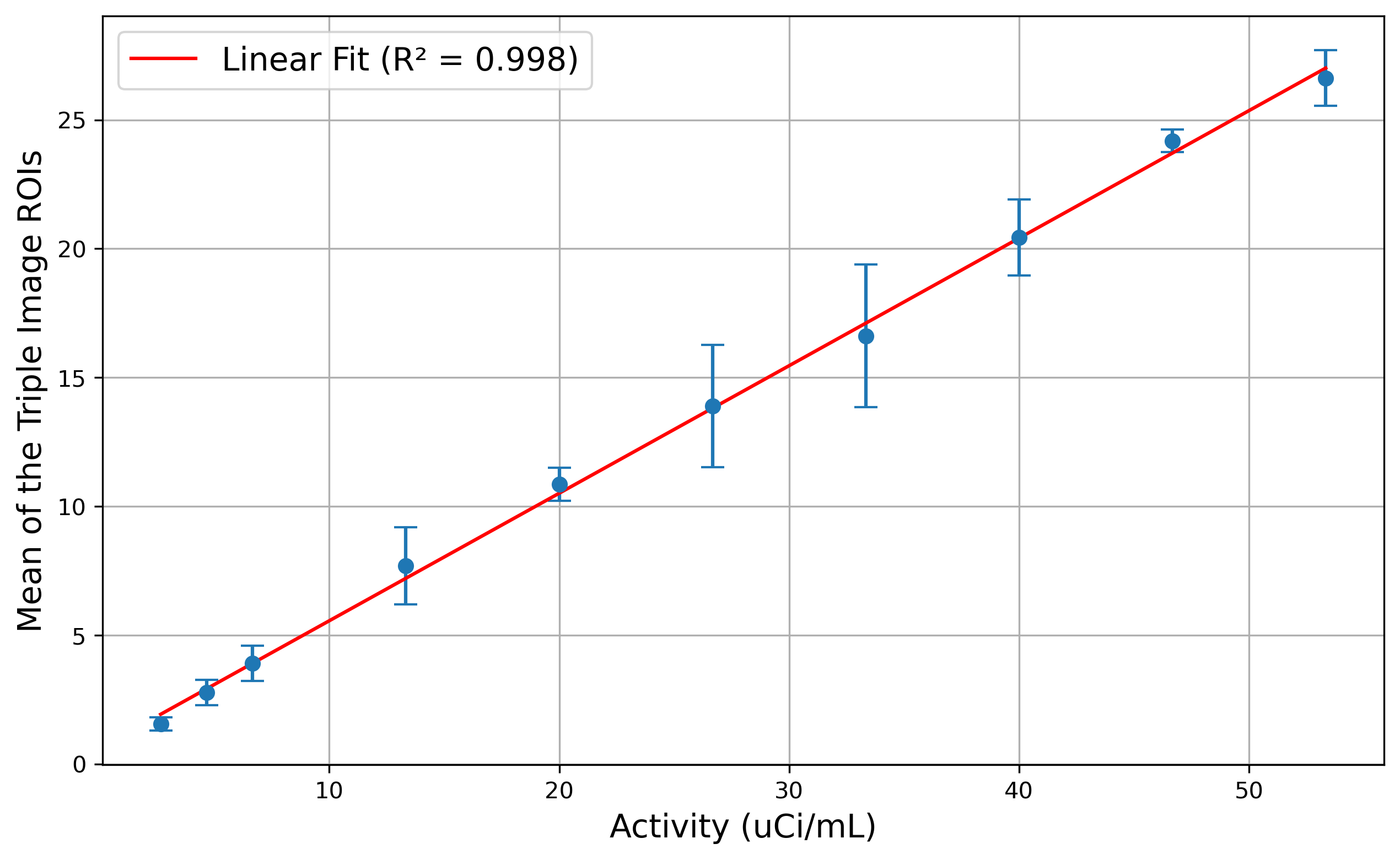}
        \subcaption{}
    \end{minipage}
    \caption{Comparison of mean image ROI values (with standard deviation) of 10 tubes for double (a) and triple (b) images vs. activity of the tube.}
    \label{fig:co55_decay_comparison}
\end{figure}

We noticed that the number of detected double and triple coincidences increased linearly with the total activity in the FOV with $R^2 = 0.9996$ for double coincidences and $R^2 = 0.9972$ for triple coincidences.
We note that the ratio of detected double coincidences to detected triple coincidence remained relatively consistent over the total activity range of 40 $\mu$Ci - 800 $\mu$Ci and stays within a 5\% range of the average ratio (11.95) as seen in Fig. \ref{fig:double_triple_event}.
Following from the total event counts, there is a linear relationship between mean of ROIs and activity concentration for both double and triple images with $R^2 = 0.998$ (Fig. \ref{fig:co55_decay_comparison}).

\subsubsection{Hollow Sphere Phantom}
\begin{figure}[ht]
    \centering
    \begin{minipage}[b]{0.45\linewidth}
        \centering
        \includegraphics[width=\linewidth]{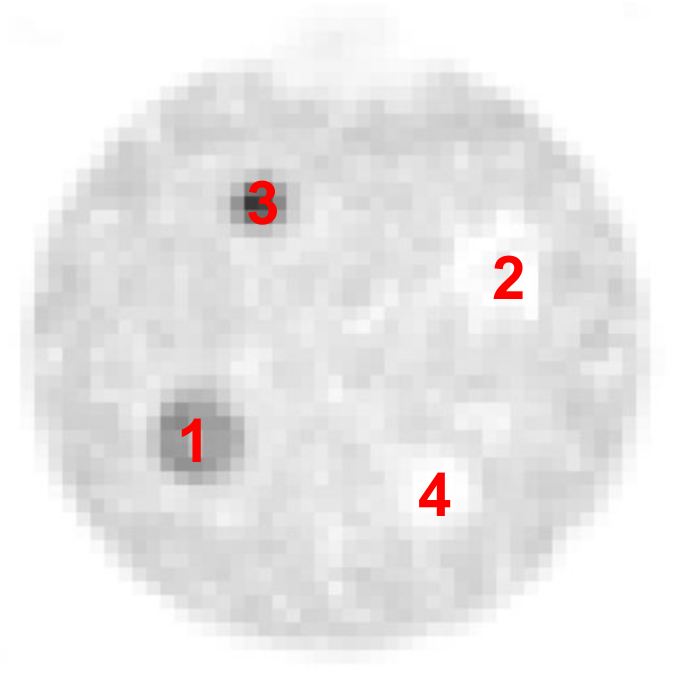}
        \subcaption{FDG/F-18 separated image}
    \end{minipage}
    \hspace{0.05\linewidth} %
    \begin{minipage}[b]{0.45\linewidth}
        \centering
        \includegraphics[width=\linewidth]{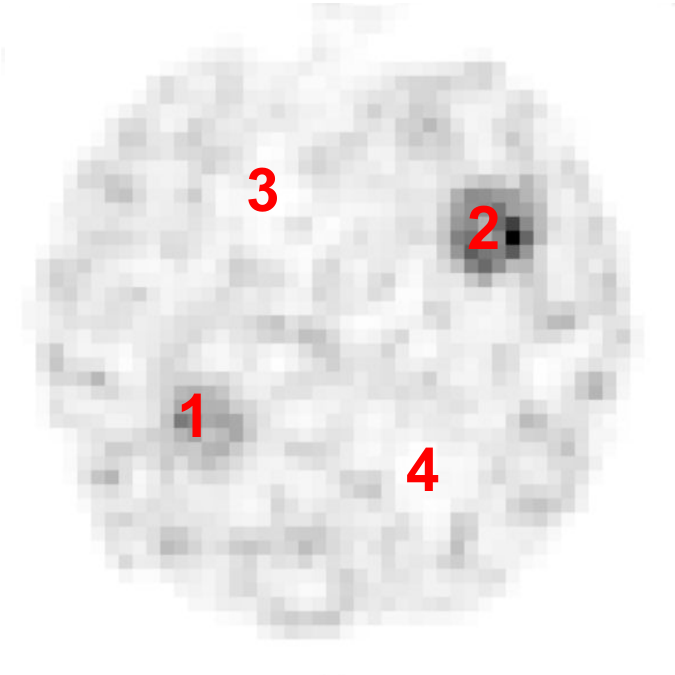}
        \subcaption{Co-55 separated image}
    \end{minipage}

    \caption{Transaxial image slices through the spheres in the hollow sphere phantom with spheres labeled: 1) 1:1 F-18:Co-55, 2) Co-55 only, 3) F-18 only, and 4) water.}
    \label{fig:sphere_im}
\end{figure}

Transaxial image slices through the spheres in the hollow sphere phantom images for F-18 and Co-55 are shown in Fig. \ref{fig:sphere_im} with sphere number labelled. Notice that sphere 1 (F-18 + Co-55) shows up as hot in both F-18 and Co-55 images due to it being filled with a 1:1 mixture of F-18 and Co-55. Sphere 2 (Co-55 only) shows up as cold in F-18 image and hot in Co-55 image as one would expect. For sphere 3 (F-18 only), it is hot in F-18 image and cold in Co-55 image. As sphere 4 is filled with water, it shows up cold in both images.  

\subsection{Animal Scans}
\begin{figure}[ht]
    \centering
    \begin{minipage}[b]{0.45\linewidth}
        \centering
        \includegraphics[width=\linewidth]{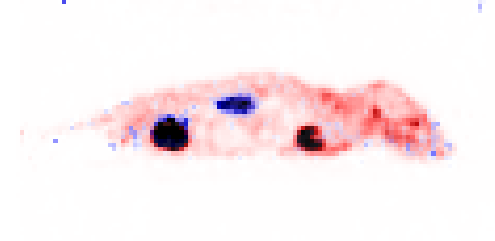}
        \subcaption{Mouse injected with two tracers.}
        \label{fig:dual_tracer}
    \end{minipage}
    \hspace{0.05\linewidth} %
    \begin{minipage}[b]{0.45\linewidth}
        \centering
        \includegraphics[width=\linewidth]{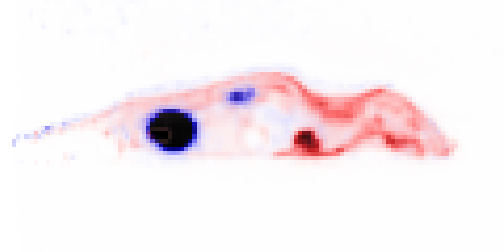}
        \subcaption{Single tracer mice images overlaid}
        \label{fig:single_tracer}
    \end{minipage}

    \caption{Comparison of (a) dual tracer mouse image and (b) overlaid single tracer mice images. Red is the FDG image and blue is the Co-55-labeled antibody image.}
    \label{fig:mice}
\end{figure}

The two mice injected with single tracers are overlaid with the FDG-only mouse scan in red and the Co-55-antibody-only mouse scan in blue (Fig. \ref{fig:single_tracer}). This is compared to the mouse injected with both tracers, where the FDG image is shown in red and the Co-55-antibody image is shown in blue in the sagittal image slice presented in Fig. \ref{fig:mice}.

\section{Discussion}
\subsection{Phantom Studies}
\subsubsection{Scan of Tubes of Uniform Concentration}

The observation of 11.38 double coincidences for every triple coincidence aligns with expectations, as 75\% of Co-55 positron emissions are accompanied by a prompt gamma. Prior studies report a measured sensitivity of 0.068 for a point source at the center of the field of view (FOV) for the Inveon system \cite{visser_spatial_2009}. Assuming the Inveon has 5\% sensitivity to double coincidences and 1\% to triple coincidences, the probability of detecting 2 out of 3 photons from a triple coincidence is given by \( P(2 \text{ out of } 3) = \binom{3}{2} \times 0.05 \times (1 - 0.22) \), where 0.22 represents the assumed singles detection rate. This yields an expected double-to-triple event ratio of \( \text{Ratio} = \frac{(25\% \times 5\% + 75\% \times 11.7\%)}{(75\% \times 1\%)} \approx 13 \). The observed double-to-triple event ratio of 11.38 is reasonably close to the theoretical estimate of 13. The difference from the theoretical estimate may be due to a lower-than-expected sensitivity to the 931 keV prompt gamma, as this energy exceeds the upper window of 814 keV and therefore requires scattering to be detected.

 \subsubsection{Hollow Sphere Phantom}
 The ratios of the mean of hot sphere ROIs to background ROI is shown in Table \ref{tab:hot_tube_comparison}. We hypothesize the 15\% underestimation for Sphere 2 in Co-55 image was due to randoms contributing to a warmer background in the triples image. The small errors observed in both isotope images for mixed Sphere 1 and in the F-18 image for Sphere 2 suggest that the image separation method performs reasonably well.

 \begin{table}[h]
\centering
\caption{Comparison of Hot Tube to Background Ratio, Ground Truth (GT), and Percent Error}
\label{tab:hot_tube_comparison}
\resizebox{.5 \textwidth}{!}{
\begin{tabular}{lccc}
\toprule
\textbf{Hot Tube to Background Ratio} & \textbf{Ratio} & \textbf{GT Value} & \textbf{Percent Error (\%)} \\
\midrule
Sphere 1: F-18   & 3.33  & 3  & 10.85 \\
Sphere 1: Co-55  & 2.75  & 3  & 8.33  \\
Sphere 2: F-18  & 0.02  & 0  & -  \\
Sphere 2: Co-55 & 5.08  & 6  & 15.34 \\
Sphere 3: F-18  & 5.84  & 6  & 2.71  \\
Sphere 3: Co-55  & 0.15  & 0  & -   \\
Sphere 4: F-18  & 0.17  & 0  & -   \\
Sphere 4: Co-55 & 0.42  & 0  & -   \\
\bottomrule
\end{tabular}
}
\end{table}

 \subsection{Animal Scans}
 The Co-55 image in the dual-tracer mouse (Fig. \ref{fig:dual_tracer}) is noisy due to lower statistics of triple coincidences.
The accumulation of FDG in the body is consistent with previous literature \cite{pauwels_fdg_1998}, showing high uptake in the heart, brain, and bladder. The accumulation of free Co-55 in the kidneys is not described in the literature thus leading to our assumption of unstable Co-55 antibody complex in the final product \cite{mitran_affibodymediated_2019, rosestedt_evaluation_2017}. However, despite the unexpected biodistribution of the Co-55-labeled antibody, this study still enables us to evaluate the unmixing methodology for the dual-tracer study. The unmixed dual-tracer image (Fig. \ref{fig:dual_tracer}) clearly shows separated tracer distributions that are similar to those observed in the overlaid single-tracer mouse distributions (Fig. \ref{fig:single_tracer}). The small differences between the two sets of distributions can be attributed to the use of three different mice. The ability to recover similar tracer distributions in the dual-tracer mouse as in the single-tracer mice indicates that the unmixing methodology for simultaneous Co-55 and FDG imaging was successful.

\section{Conclusion}
We successfully demonstrated mPET with Co-55 and F-18 signals using the Inveon preclinical scanner. This approach allows for the evaluation of two (or potentially more) biomarkers, which is crucial for characterizing cancer and assessing its response to novel treatments, such as cancer immunotherapy. Such treatments require a multi-parametric assessment to provide a more comprehensive understanding of treatment efficacy.

\section{Compliance with ethical standards}
\label{sec:ethics}
All experiments in this work was conducted under Stanford University's Institutional Animal Care and Use Committee (protocol 21585) in strict compliance with ethical standards. Radioactive work was conducted under the Controlled Radiation Authorizations (CRAs) approved by Stanford Health Physics.

\section{Acknowledgments}
\label{sec:acknowledgments}

This work was supported by Dr. Ralph \& Marian Falk Medical Research Trust (SPO 268891), Stanford Bio-X Interdisciplinary Initiatives Seed Grants Program (IIP), and Stanford c-ShaRP Program. The Inveon scanner was procured in part with NIH grant 1S10OD018130-01. The UAB cyclotron facility is a member of the Department of Energy University Isotope Network and is supported through DESC0021269 (PI: Lapi). The authors would like to thank F. Habte with the Stanford Center for Innovation in In vivo Imaging and H. Redman with Stanford Health Physics. 

\bibliographystyle{IEEEbib}
\bibliography{strings,references,ref}

\end{document}